\documentclass[prl,aps,twocolumn,showpacs,amsmath,amssymb,floatfix]{revtex4}
\usepackage{times} 
\usepackage{graphicx}

\usepackage{xcolor}
\newcommand{\changed}[1]{#1}
\usepackage[normalem]{ulem}

\begin{document}
\title{Energy Dissipation in Driven Granular Matter in the Absence of Gravity}
\author{Achim Sack}
\author{Michael Heckel}
\author{Jonathan E. Kollmer}
\author{Fabian Zimber}
\author{Thorsten P\"oschel}
\affiliation{Institut f\"ur Multiskalensimulation, Friedrich-Alexander-Universit\"at Erlangen-N\"urnberg, Germany}
\date{\today}

\begin{abstract}
  We experimentally investigate the energy dissipation rate in
  sinusoidally driven boxes which are partly filled by granular
  material under conditions of weightlessness. We identify two
  different modes of granular dynamics, depending on the amplitude of
  driving, $A$. For intense forcing, $A>A_0$, the material is found in
  the {\em collect-and-collide} regime where the center of mass of the
  granulate moves synchronously with the driven container while for
  weak forcing, $A<A_0$, the granular material exhibits gas-like
  behavior. Both regimes correspond to different dissipation
  mechanisms, leading to different scaling with amplitude and
  frequency of the excitation and with the mass of the granulate. For
  the collect-and-collide regime, we explain the dependence on
  frequency and amplitude of the excitation by means of an effective
  one-particle model. For both regimes, 
  the results may be collapsed to a single curve
  characterizing the physics of granular dampers.
\end{abstract}


\maketitle

\emph{Introduction.}  When containers filled with granular material
are subjected to vibrational motion, a number of interesting phenomena
is observed.  The most prominent among them are self-organized
convection flows and various segregation effects which are reported in
a large body of literature,
e.g. \cite{Knight:1993,Gallas:1992,AransonTsimring:2009}. Most, if not
all, of these effects found in driven granular systems are influenced
by gravity, therefore, to study these effects in absence of gravity,
experiments have been performed under conditions of weightlessness in
parabolic flights, drop towers, and sounding rockets. Examples for
such investigations concern shear flow \cite{Murdoch:2013}, cooling
and clustering in dilute systems
\cite{Falcon:1999,Tatsumi:2009,Grasselli:2009,Harth:2013}, violations of the
energy equipartition \cite{Chen:2012,Hou:2008,Leconte:2007}, the
propagation of sound \cite{Zeng:2007}
\changed{ and segregation \cite{Guettler:2013}}.

In this letter, we experimentally address the mechanisms of energy
dissipation in vibrationally driven granular systems in the absence of
gravity.

Dissipation mechanisms belong to the fundamental properties of matter
and are of interest {\em per se}. But besides scientific curiosity,
the investigation of dissipation in granular systems is of
technological interest to dampen unwanted vibration using devices which are
referred to as {\em granular dampers} which are containers or
cavities filled with granular material. When the container is
subjected to oscillatory motion, the grains inside the cavity collide
inelastically with one another and with the confinement and, thus,
dissipate mechanical energy of the vibration. In contrast to
conventional dashpot dampers, granular dampers do not rely on a fixed
anchor as an impulse reservoir. They show only a minute dependence on
temperature, and since such dampers can be sealed off hermetically,
they are predestined for long term use in harsh environment with
extreme temperatures and/or high pressures. For example, using
granular dampers, turbine blades can be kept from oscillating and in
medical tools where sterilization is mandatory they can dampen the
vibration of handles
\cite{Kielb:1999,Heckel:2012,Norcross:1967,Xia:2011}.

Extensive research has been performed on understanding the behavior of
granular dampers in specific applications and to determine the
influence of material and construction parameters like the size of the
container or cavity, the number of particles, the mass of the filling,
and the clearance,
e.g. \cite{Sadek:1970a,Sadek:1970b,Saluena:1999,Yang:2003,Bai:2009b,Sanchez:2011,Sanchez:2012,Sanchez:2013,Cui:2011}.
Recent simulations show the flow of the granular material inside such
a cavity under normal gravity for a vertically shaken system
\cite{Sanchez:2012,Sanchez:2013}.

From all of these investigations it may be concluded that the
dissipation properties of a granular damper depend in a non-trivial
way on many variables like the amplitude of the oscillation, the
clearance, the frequency as well as the mass and type of particles
\cite{Marhadi:2005,Ham:2012}. By now, however, these dependencies are
poorly understood.

Besides the lack of understanding the physics of such systems, from a
technological point of view there is a need for a master design curve,
in order to be able to {\em predict} the damping properties of a
granular damper in a certain application.

As shown in the comprehensive work by Yang \cite{Yang:2003}
\changed{for vertical excitation in gravity, 
only for strong forcing (high excitation energy), it is possible to collapse the dissipated power as
a function of the amplitude for various frequencies and, thus, to
deduce an empirical master design curve.}  For less intense driving it
was shown by means of numerical simulations \cite{Saluena:1998} that
to a large extend the dissipative properties of granular dampers are
influenced by gravity, in particular when gravity exceeds the
acceleration due to the external forcing.

\changed{To isolate the response of driven granular systems to external excitation from the disturbing effect of gravity, in this letter for the first time we address the energy dissipation of granular dampers under conditions of
weightlessness during parabolic flights.} We explain the dependence on
frequency and amplitude of the excitation by means of an effective
one-particle model. Without using any adjustable parameter the results
may be collapsed to a single curve characterizing the physics of
granular dampers.

\begin{figure}[ht]
  \centerline{\includegraphics[clip,width=\columnwidth]{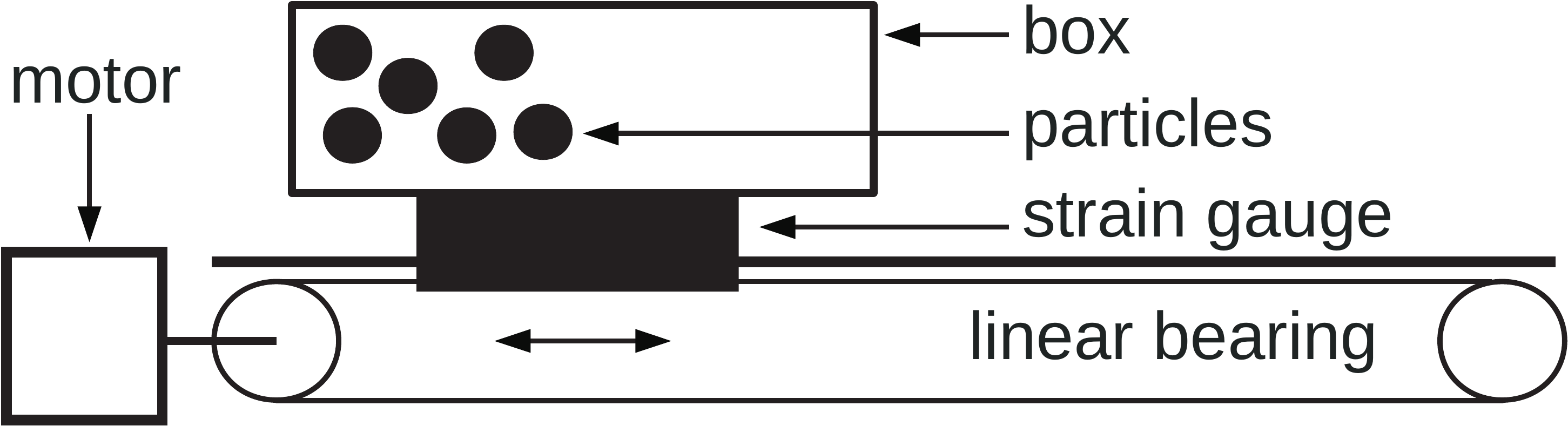}}
\caption{\label{fig:experimentaldesign}
  Sketch of the experiment. For explanation see the text.}
\end{figure}

\emph{Experiment.} The experimental setup is sketched in
Fig. \ref{fig:experimentaldesign}. The sample box is mounted on a
strain gauge which in turn is attached to a carrier moving on a linear
bearing. A gear belt connects the carrier with a computer controlled
stepper motor which drives the carrier to perform sinusoidal
oscillations of adjustable angular frequency, $\omega$, and amplitude,
$A$. The time-dependent position of the container was measured using
Hall-effect based position encoders with a resolution of 20$\,\mu$m
and 10\,kHz sample rate to check that the deviation of the container's
motion from the set sinusoidal oscillation is negligibly small\changed{, typically
the spurious free dynamic range is 56 dB.} The experiment was monitored
by a high-speed camera at a frame rate of 240 fps.
The entire setup was built up twice such that the containers move
in opposite direction to cancel the vibrations transmitted to the
external mounting structure.

The strain gauge delivered signals \changed{proportional to} the forces in the direction of
driving, and the two directions perpendicular to it. Only the first
one is relevant for our experiment; we checked that the forces in the
other directions are negligibly small as compared to it, \changed{i.e., the 
side walls confine the granular material while the main transfer
of momentum occurs parallel to the direction of motion.}

The samples consist of polycarbonate boxes (wall thickness 4~mm)
partially filled by different amounts of steel beads (diameter 4\,mm,
material density $7.8$ g/cm$^3$, Young's modulus $203.5$ GPa). The
number and the total mass of particles within each sample
is given by $N$
and $m$, respectively.
The clearance, $L_g$, is the difference between box length and
the thickness of the packed layer of particles in the box. It can be
obtained by computing the volume occupied by particles in random close
packing \cite{Aste:2000} at volume fraction 64 \%.
Table \ref{tab:parameters} summarizes the characteristics of our samples.
\begin{table}[h]
  \begin{center}
    \caption{\label{tab:parameters} 
      Table of Samples. 
}
    \begin{tabular}{@{\extracolsep{\fill}}|c|c|c|c|c|}
      \hline
      Sample No. &  Box L$\times$W$\times$H (mm$^3$) & $m$ (g) & $N$ & $L_g$ (mm)\\
      \hline
      $1$ & $100\times50\times50$ & $126.3$ & $473$ & $89.4$ \\
      \hline
      $2$ & $50\times50\times50$ & $135.3$ & $507$ & $38.7$ \\
      \hline
      $3$ & $50\times50\times50$ & $71.0$ & $266$ & $44.1$ \\
      \hline
      $4$ & $100\times50\times50$ & $63.8$ & $239$ & $94.7$ \\
      \hline
    \end{tabular}
  \end{center}
\end{table}

To exclude the influence of gravity, the experiment was performed
during a parabolic flight allowing for stable microgravity condition
$(0 \pm 0.05)$ g for time intervals of about $22$ seconds which
determines the duration of each single measurement, where amplitude
and frequency of the excitation were fixed. The data resulting from
the strain gauges and the position sensors were sampled simultaneously
at a rate of $10\,$kHz and stored for later evaluation of the dissipated
energy. About four seconds after the onset of microgravity the
experiment had entered the stationary state which could be identified
by both the rate of dissipation deduced from the measurement of the
driving force and the recordings of the high-speed camera. For the
results reported here we use only the data obtained in the
stationary state.

\begin{figure}[ht]
  \centerline{\includegraphics[clip,width=\columnwidth]{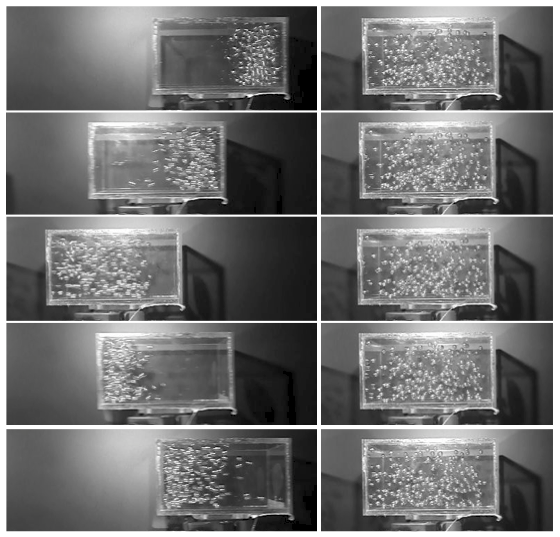}}
  \caption{\label{fig:longbox} Snapshots from the high-speed video
    recordings of sample 4 illustrating the two distinct regimes of
    excitation: collect-and-collide regime at $A=50~\text{mm}$ (left
    column) and a gas like state at $A=2.5~\text{mm}$ (right
    column). Each column shows the box at a phase range from $0$ (top)
    to $\pi$ (bottom).  }
\end{figure}

\emph{Regimes of Dynamical Behavior.} Analyzing the high-speed video
recordings we can identify two different regimes of dynamical
behavior, see Fig. \ref{fig:longbox}. For large amplitudes of the
vibration, the damper operates in the collect-and-collide regime, that
is, during the inward stroke all the material is ``collected'' and
accumulates as a relatively densely packed layer at the wall of the
container. After passing the phase of maximal velocity, the box
decelerates and the layer of particles leaves the wall
collectively. When the bulk of particles impacts the opposite wall of
the container, a large part of the kinetic energy is dissipated by
inelastic collisions. The amount of energy dissipated depends on the
relative velocity between the particles and the wall at the time of
the impact, determined by the amplitude and frequency of the vibration
and the filling ratio of the container. The collect-and-collide regime
was theoretically and experimentally investigated in
\cite{Bannerman:2011} and identified as the regime of most efficient
damping. It was confirmed also by numerical MD-simulations
\cite{Opsomer:2011,Opsomer:2011a} and identified as one out of four
different regimes of dynamical behavior of vibrated granulate in
microgravity.

For small amplitudes we observe a gaseous state where only a small
fraction of the particles interact with the oscillating walls during
one oscillation period.  In the gaseous state, the collisions of the
particles with the driving walls are just sufficient to balance the
energy loss according to dissipative particle-particle collisions in
the bulk of the material. Here, the dissipation rate is smaller than
in the collect-and-collide regime \cite{Bannerman:2011}.

\emph{Energy Dissipation Rate.}  To obtain the energy dissipated by
the granulate during one period, $T\equiv 2\pi/\omega$, of the sinusoidal driving, $x=A\sin(\omega t)$, we integrate the product of the measured force, $F(t)$, and velocity,
$\dot{x}(t)$, over one period of oscillation:
\begin{equation}
  E_\text{diss}\equiv\int\limits_T \dot{x}(t)F(t)  \, \mathrm{d} t.
\end{equation}

\changed{The maximum energy that can be dissipated during one cycle in the system is given by:
\begin{equation}
  E_\text{max}=4 m A^2 \omega^2. 
\end{equation}
This is the case if all particles collide inelastically with the wall at 
maximum relative velocity. In the following, $E_{\text{max}}$ is used for normalization.}

\begin{figure}[ht]
  \centerline{\includegraphics[clip,width=\columnwidth]{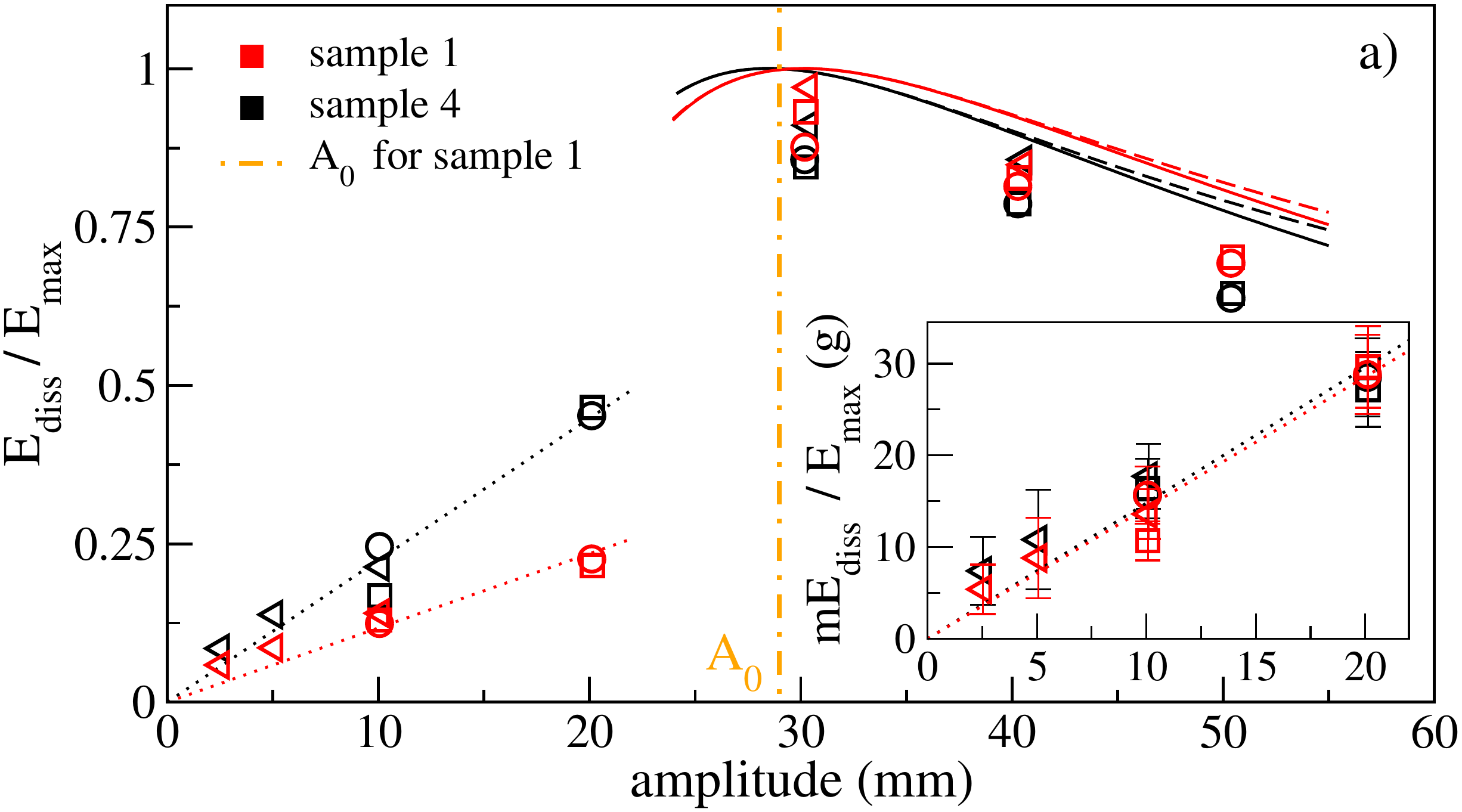}}
  \centerline{\includegraphics[clip,width=\columnwidth]{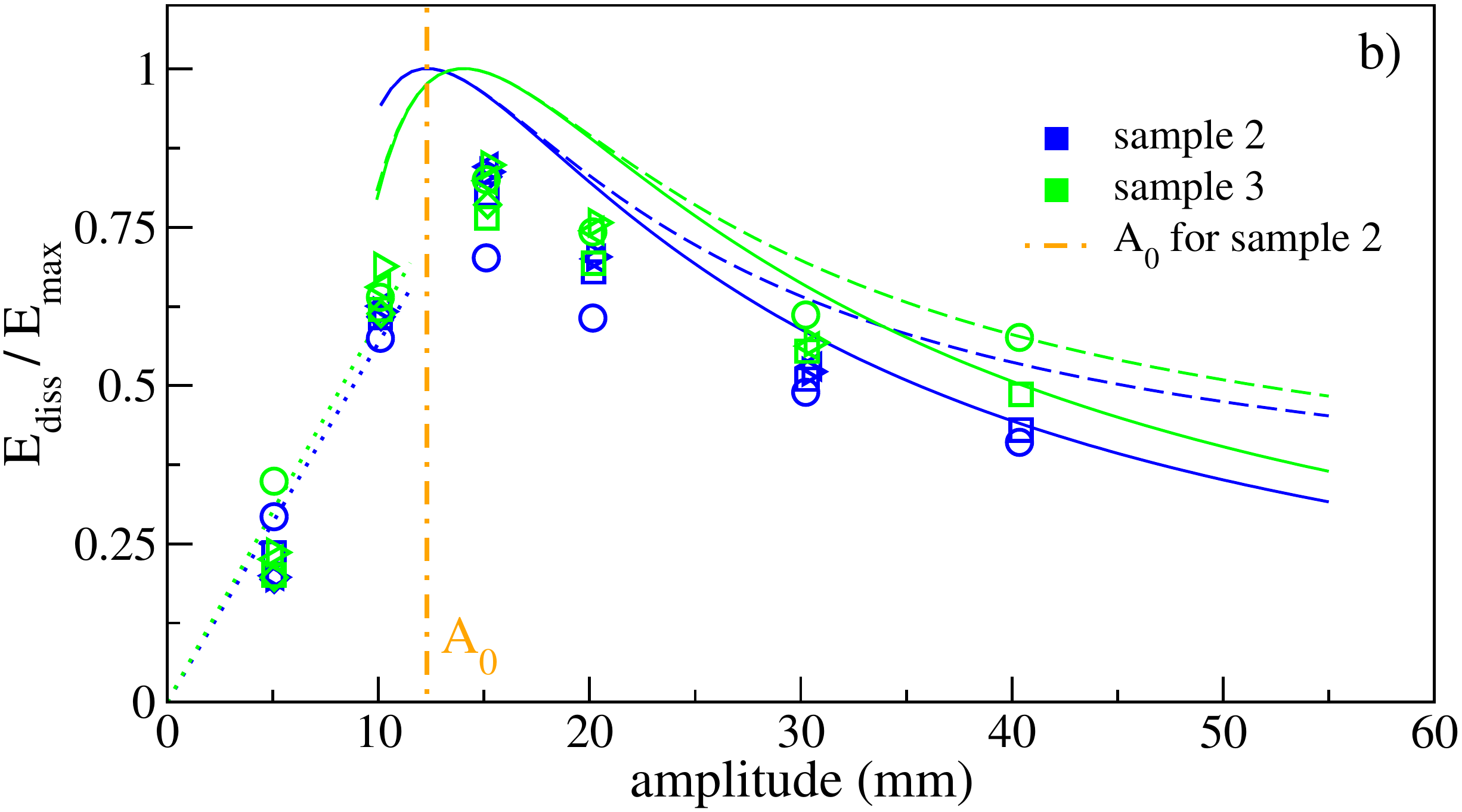}}
  \caption{\label{fig:impactmodel} (color online) Normalized
    dissipated energy per period of external vibration.  Symbols:
    Experimental data. \changed{$\bigcirc: 1$\,Hz, $\Box: 2$\,Hz; $\Diamond: 3\,$Hz, $\triangleleft: 4$\,Hz; $\triangleright: 5$\,Hz.}
    Lines: Solution of the
    impact model, valid for $A > A_0$ (solid -- numerical, dashed --
    analytical, Eq. \eqref{eq:rpd}).  Dotted lines: Dissipation rate
    for the gaseous regime ($A < A_0$, see Eq. \eqref{eq:Egas}). Inset: same data (only gas
    regime) but normalized to \changed{$E_{\text{max}}/m$ (see
    Eq. \eqref{eq:Egas1})}. The error bars for the gas regime are shown
    in the inset. For all other measurements, the errors are about the
    size of the symbols. The threshold amplitude, $A_0$, (vertical
    lines) obtained from the model (see Eq. \eqref{eq:A0}) agrees with
    the experimental data.  }
\end{figure}

We measured the dissipated energy per period for the following ranges
of frequency and amplitude: Sample $1$ and $4$ were shaken at $1\,$Hz, $2\,$Hz,
$4\,$Hz while samples $2$ and $3$ were shaken from $1\,$Hz to $5\,$Hz in $1\,$Hz
increments. For each setup, Fig. \ref{fig:impactmodel} shows $E_\text{diss}/E_\text{max}$ \changed{versus} the amplitude of the oscillation.

Let us first consider the gas-like state observed for small amplitude,
$A<A_0$. In this regime, we 
\changed{expect the dissipated energy to be proportional to the number of particles colliding with the wall.
If we assume a monodisperse system with homogeneous density this number is determined by the the volume swept by the container's wall. We further assume the characteristic velocity of the particles to scale with the velocity of driving, $A\omega$, and the particles hitting the wall at
random phases, due to their disordered motion and arrive at:}
\changed{\begin{equation} 
  \label{eq:Egas}
  E_\text{diss}^g  \propto m \frac{A^3\omega^2}{L} =  \frac{A}{4L} E_\text{max}\,,
\end{equation}
}

Note that particle-particle collisions in the bulk of the
material contribute only indirectly to $E_\text{diss}^g$ since such
collisions do not transfer momentum to the container.

Equation \eqref{eq:Egas} was developed under the assumption of a homogenous density distribution.
This however may not always hold true.
Unlike molecular
gases, heated granular gases are not homogeneous but density increases
in a non-linear way with distance from the driving wall
\cite{Grossman:1997} to form regions of enhanced density (clusters)
far away from the wall. \changed{Following the arguments of \cite{Meerson:2004}
the number of particle 
wall collision depends only weakly on the total mass of particles in the system. 
Consequently, for the limit of no dependence on the total mass} we may write
\changed{\begin{equation} 
  \label{eq:Egas1}
  \frac{E_\text{diss}^g}{E_\text{max}} \propto \frac{A}{4Lm} \,,
\end{equation}}
shown in the inset of Fig. \ref{fig:impactmodel}a. 

\changed{For the cases described by Eqs. (\ref{eq:Egas},\ref{eq:Egas1})},
the experimental data shown in Fig. \ref{fig:impactmodel} collapse despite the fact that the data
points shown for a certain amplitude correspond to different
frequencies of driving. That is, the dissipation rate is independent
of frequency in agreement with the scalings,
\changed{Eq. (\ref{eq:Egas1}) for Fig. \ref{fig:impactmodel}\,a) and
Eq. (\ref{eq:Egas}) for Fig. \ref{fig:impactmodel}\,b).}

For large amplitudes, $A>A_0$, when the granulate is in the
collect-and-collide regime, the experimental data collapse when scaled
with $E_\text{max}$. Moreover, also here the scaled data
belonging to the same amplitude are independent of the frequency. To
quantitatively explain the data shown in Fig. \ref{fig:impactmodel}
and to understand its independence of frequency for $A>A_0$, we take a
closer look at the dynamics of the collect-and-collide regime, where
nearly all particles are collected by the inward stroke and leave the
wall collectively at $t=0$ with velocity $v_\text{particle}=A\omega$. \changed{We define $t_c$ as the time
when the particles collide with the opposing wall where they will adopt its
instantaneous velocity $v_\text{wall} = A\omega \cos(\omega t_c)$.
Note that this corresponds to the motion of a
quasi-particle interacting perfectly inelastically with the
container walls. For a justification of this model see
\cite{Bannerman:2011, Sanchez:2012, Luding:1994}.} The amount of kinetic energy lost per
period depends on the difference of the velocity of the quasi-particle and the wall: 
\changed{\begin{equation}
  \label{eq:Ediss}
  E_\text{diss}^\text{cc}= m(v_\text{particle}-v_\text{wall})^2\,.
\end{equation}}
Expressed in terms of $t_c$ and $E_\text{max}$ we obtain:
\begin{equation}
  \label{eq:Ediss1}
  E_\text{diss}^\text{cc}  = \frac{1}{4} [1-\cos(\omega t_c)]^2 E_\text{max}\,.
\end{equation}

The time $t_c$ is implicitly given by the distance the bulk of
particles has traveled and the harmonic motion of the box,
\begin{equation}
  \label{eq:tc_implicit}
  v_\text{wall}t_c =  A\omega t_c = A \sin(\omega t_c) + L_g \textrm{\,.} 
\end{equation}
Equation \eqref{eq:tc_implicit} can be solved numerically for $\omega
t_c$ to obtain the dissipated energy per period,
$E_\text{diss}^\text{cc}/E_\text{max}$, via Eq. \eqref{eq:Ediss1} (see
solid lines in Fig. \ref{fig:impactmodel}).

Alternatively, we can obtain an approximate value for $\omega t_c$ by
a first order expansion of Eq. \eqref{eq:tc_implicit} around $\omega
t_c = \pi$: \begin{equation}
  \label{eq:tc_analytic}
  \omega t_{c} \approx \frac{\pi}{2} +\frac{L_g}{2A} \,.
\end{equation}
Inserting this solution into Eq. \eqref{eq:Ediss1} we find
\begin{equation} 
  \label{eq:rpd}
    \changed{\frac{E_\text{diss}^\text{cc}}{ E_\text{max}} \approx \frac{1}{4} \left[ 1-\sin \left( \frac{L_g}{2A} ~\right) \right]^2 \, .}
\end{equation}
Figure \ref{fig:impactmodel} compares the relative dissipated energy
per oscillation period as obtained in experiments (symbols) with the
numerical solution of the collect-and-collide model, Eqs.
(\ref{eq:Ediss},\ref{eq:tc_implicit}), (solid lines) and its
approximate analytical solution, Eq. \eqref{eq:rpd} (dashed lines),
resulting in good agreement. Also in agreement with the experimental
data, both Eq. \eqref{eq:rpd} and Eq. \eqref{eq:Ediss1} with $\omega
t_c$ from Eq. \eqref{eq:tc_implicit} and Eqs. (\ref{eq:Egas},\ref{eq:Egas1}) are independent of the
frequency, $\omega$, which explains the collapse of the data for
different frequencies. Note that the model, Eqs.
(\ref{eq:Ediss1},\ref{eq:tc_implicit}), and the approximate solution,
Eq. \eqref{eq:rpd}, do not contain any adjustable parameters.

Furthermore, this model provides an explanation of the threshold $A_0$
separating the gaseous state from the collect-and-collide regime: The
bulk of particles leaves the wall on its inward stroke at time $t=0$
(instant of maximal velocity). If \changed{it arrives} at the opposite wall at
a time where the wall is accelerating away from \changed{it}, that is $\pi <
\omega t_c < 2\pi$, \changed{it} will not get collected by the wall but mainly
scattered. This scattering inevitably leads to desynchronization of
collective particle motion and the collect-and-collide mode breaks
down. The threshold can be obtained from Eq. \eqref{eq:tc_analytic}
with $\omega t_c = \pi$: \begin{equation}
  \label{eq:A0}
  A_0 = \frac{L_g}{\pi}\,.
\end{equation}
Interestingly, at the edge of stability of the collect-and-collide
regime, $A=A_0$, that is, $\omega t_c = \pi$, from Eq.
\eqref{eq:Ediss1} we obtain maximal efficiency in damping. The
threshold amplitude, $A_0$, obtained from the model agrees well with
the experimental data, see Fig. \ref{fig:impactmodel} (vertical
lines).

\emph{Conclusion.}
\changed{On Earth, the influence of gravity on granular dynamics can only be neglected
for intense driving, where $A^2\omega^2 \gg g$. To study the response of granular dampers to external excitation isolated from the disturbing effect of gravity, we}
investigated the energy dissipation of granular
matter in the absence of gravity when subjected to sinusoidal motion,
$x(t)=A\sin(\omega t)$.

Depending on the amplitude of the vibration, we
observe two qualitatively different modes: For small
amplitude, $A <A_0 \approx L_g/\pi$, the
granular material behaves gas-like while for larger amplitudes,
$A>A_0$, we observe a {\em collect-and-collide} regime where the
center of mass of the granulate moves synchronously with the driven
container. In this regime, we describe the system by an effective
one-particle model, Eq. \eqref{eq:Ediss1}, which can be solved
numerically or using an analytical approximation. \changed{Despite of the model's simplicity, we find remarkably good agreement with the experimental data.}

In the gas regime, the energy dissipation rate increases linearly with
the amplitude $A$. For large clearance, $L_g$, the energy dissipation
rate is almost independent of the mass of the granulate in agreement
with the known density instability in granular gases, Eq.
\eqref{eq:Egas1}. Here an addition of particles hardly increases the
density in the vicinity of the heated wall \cite{Meerson:2004}. For
smaller clearance, the high-density region is no longer far away from
the driving wall such that the energy dissipation rate increases
linearly with the mass, Eq. \eqref{eq:Egas1}

In both regimes, gas-like and collect-and-collide, the energy
dissipation rate was found independent of the frequency, in agreement
with our model description, Eqs. (\ref{eq:Egas1},\ref{eq:Ediss1}).

With appropriate scaling, for a fixed container geometry all
measurements coincide along one universal ``master curve'', independent
of filling fraction, mass of the
granulate, particle properties, particle number, and the parameters of
vibration. Based on the theory of granular gases for the gas state and
on the effective one-particle model for the collect-and-collide state,
we derived a model description without any fitting parameters which
describes the experiment up to a good agreement. Although not measured, from the model it may be 
\changed{expected that the dissipative properties are invariant of the type of granulate, the particle diameter and the exact cross section of the cavity}.

\emph{Acknowledgments.} We thank the European Space Agency (ESA) for
funding the parabolic flight campaign and the German Science
Foundation (DFG) for funding through the Cluster of Excellence
``Engineering of Advanced Materials''.


\end{document}